\documentclass[10pt]{article}
\usepackage{color,amsmath,amssymb,epsfig,float,graphics,fullpage}
\usepackage[dvips]{hyperref}
\title{Mathematical modeling of the GnRH pulse and surge generator}

\author{Fr\'ed\'erique Cl\'ement\thanks{Unit\'e de recherche INRIA Rocquencourt,
Domaine de Voluceau, Rocquencourt BP 105, 78153 Le Chesnay Cedex, France. {\tt frederique.clement@inria.fr}}
        \and Jean-Pierre Fran\c{c}oise\thanks{Laboratoire Jacques Louis Lions, 
Universit\'e P.-M. Curie, Paris VI, UMR 7598 CNRS,
175 Rue du Chevaleret, 75013 Paris, France. {\tt jpf@ccr.jussieu.fr}}}
\newcommand{\ds}{\displaystyle}
\newcommand{\eref}[1]{(\ref{#1})}
\begin{document}

\maketitle
\topmargin=0cm
\paperwidth=21.9cm
\paperheight=29cm

\begin{abstract}
We propose a mathematical model allowing for the alternating pulse and surge pattern of GnRH (Gonadotropin Releasing Hormone) secretion. The model is based on the coupling between two systems running on different time scales. The faster system corresponds to the average activity of GnRH neurons, while the slower one corresponds to the average activity of regulatory neurons. The analysis of the slow/fast dynamics exhibited within and between both systems allows to explain the different patterns (slow oscillations, fast oscillations and periodical surge) of GnRH secretion. Specifications on the model parameter values are derived from physiological knowledge in terms of amplitude, frequency and plateau length of oscillations. The behavior of the model is finally illustrated by numerical simulations reproducing natural ovarian cycles and either direct or indirect actions of ovarian steroids on GnRH secretion.
\end{abstract}


\pagestyle{myheadings}
\pagestyle{plain}

\maketitle
\section{Introduction}
\subsection{Endocrine background}
The reproductive function involves tightly and finely controlled processes. The reproductive axis, usually called the gonadotrope axis, includes the hypothalamus, within the central nervous system, the pituitary gland and the gonads (ovaries in females, testes in males). 
Specific hypothalamic neurons secrete the gonadotropin releasing hormone (GnRH) in a pulsatile manner. The pulsatile GnRH secretion pattern ensues from the synchronization of the secretory activity
of individual GnRH neurons. The release of GnRH into the pituitarian portal blood induces the secretion of the luteinizing hormone (LH) and follicle stimulating hormone (FSH) by the pituitary gland. The changes in the frequency of GnRH pulses (between 1 pulse per hour and 1
pulse every 6 hours in the course of an ovarian cycle) has a fundamental role in the differential
control of the secretion of both gonadotropins: the secretion of LH is enhanced by higher frequencies, while that of FSH is enhanced by lower frequencies \cite{taragnat_03}. On the gonadic level, FSH and LH sustain germ cell production and hormone secretion. In turn, hormones secreted by the gonads modulate the secretion of GnRH, LH and FSH within entangled feedback loops. 

In females, the frequency of GnRH pulses is subject to the control exerted by ovarian steroids, estradiol and  progesterone. Progesterone slows down GnRH pulsatility \cite{moenter_91}, which leads to a slower frequency during the luteal phase (when progesterone levels are high) compared to the follicular phase \cite{evans_02}. On the contrary, estradiol speeds up GnRH pulse frequency, but at the expense of a decrease in pulse amplitude \cite{NPEvans1994}, so that the whole feedback effect on rate secretion is rather inhibitory (negative estradiol feedback). 

\noindent The GnRH secretion pattern alters dramatically once per ovarian cycle, resulting in the GnRH surge characterized by massive continuous release of GnRH in response to increasing levels of estradiol \cite{karsch_97} (positive estradiol feedback). The GnRH surge triggers the LH surge which is responsible for ovulation, leading to the release of fertilizable oocytes.  

\noindent The estradiol signal is conveyed to GnRH neurons by regulatory neurons (also designed as interneurons) either directly or after other neuronal relays. Transmission from interneurons calls to many different neurotransmitters  (see review in \cite{ herbison_98,smith_01}). The balance between stimulatory and inhibitory signals emanating from interneurons controls the behavior of the GnRH network \cite{herbison_98}.\\
 Rencently, a specially interesting type of regulatory neurons has been discovered. Kisspeptin neurons act on GnRH neurons via the G-protein coupled receptor GPR54 \cite{dungan_06}. They are very good candidates for relaying both positive and negative estradiol feedback, since they react to estradiol in opposite ways according to the anatomic area of the hypothalamus where they lie. \\
 
 \subsection{Model objectives}
We aim at formulating a phenomenological, data-dri\-ven model of GnRH secretion. This paper focuses on the coupling between the GnRH neuron network and the regulatory interneuron network. Each network is represented by the behavior of a single average neuron. The key point in the design of the model consists in entering as reasonable input 2 coupled systems (a slow one with a faster one)  to generate a definite sequence of events in the model output: GnRH secretion. This coupling yields a 3 time-scaled model which is able to capture, not only the cyclic transition from a pulsatile to a surge secretion pattern of GnRH, but also the increase in the pulsatility frequency  between the luteal and  follicular phases. It also separates a specific dynamical state corresponding to pulsatility resumption after the surge. Besides, parametrization of the model is subject to physiological specifications expressed as constraints on the GnRH output and allows to reproduce the direct (on the GnRH network) or indirect (via the regulatory network) effects of ovarian steroid hormones (estradiol and progesterone) on GnRH secretion.

In summary, we aim at reproducing a synthetic mathematical representation of GnRH secretion pattern, fitting available observations -in agreement with  schematic "hand-made" representations such as that  proposed by Herbison (top of Figure 4 in \cite{ herbison_98}). 

\noindent The paper is organized as follows. In section \ref{modeldesign}, we introduce the equations of the model. We comment the numerical simulations in section \ref{simul} to motivate the analysis of the bifurcations in a model-derived 2-scaled system in section \ref{bifurcation}. We finally address the question of amplitude and frequency control in section \ref{control}.
\section{Model design and analysis}\label{modeldesign}

\subsection{Coupling oscillatory neuronal dynamics}

We consider the following four-dimensional dynamical system: 
\begin{subequations}
\begin{eqnarray}
\epsilon\delta\dot{x}&=&-y+f(x)\label{eqdotx}\\
\epsilon\dot{y}&=&a_{0}x+a_{1}y+a_{2}+cX\label{eqdoty}\\
\epsilon\gamma\dot{X}&=&-Y+F(X)\label{eqdotX}\\
\dot{Y}&=&b_{0}X+b_{1}Y+b_{2}\label{eqdotY}\\
z(t)&=&\chi_{\lbrace y(t)>y_s\rbrace}\nonumber
\end{eqnarray}
\end{subequations}

\noindent Equations \eref{eqdotx} and \eref{eqdoty} correspond to a fast system representing an average GnRH neuron, while equations \eref{eqdotX} and \eref{eqdotY} correspond to the slower system representing an average regulatory neuron.
The $x,X$ variables represent the neuron electrical activities (action potential), while the $y, Y$ variables relate to ionic and secretory dynamics. The fast variables are assumed to have two stable stationary points separated by a saddle. Their bistability is accounted for by the cubic functions $f(x)$ and $F(x)$. The intrinsic dynamics of the slow variables follows a growth law of very small velocity ($a_1\! \ll\! 1$ ). In each system, the fast and slow variables feedback on each other. The coupling between both systems is mediated through the unilateral influence of the slow regulatory neurons onto the fast GnRH ones ($cX$ term in equation \eref{eqdoty}). The coupling term aggregates the global balance between inhibitory and stimulatory neuronal inputs onto the GnRH neurons. The global system exhibits 3 time scales given by $\epsilon\delta$, $\epsilon$ and $1$. Constant $\gamma$ is close to 1.\\
\noindent In many cell types, thresholding calcium dynamics is known to trigger secretion. As far as GnRH secretion is concerned, many evidence for the inducing effect  of calcium have also been gathered (see discussion in \cite{EiTerasawa07151999} for details). Hence, we associate GnRH secretion to a threshold ionic activity $y\!=\!y_s$, and finally keep as representative GnRH signal $z(t)\!=\!\chi_{\lbrace y(t)>y_s\rbrace}$.

\subsection{Mechanisms underlying the pulse to surge transition}

\subsubsection{Bifurcations in the fast GnRH system}

\begin{figure}[H]
{\centering
\includegraphics[scale=0.6]{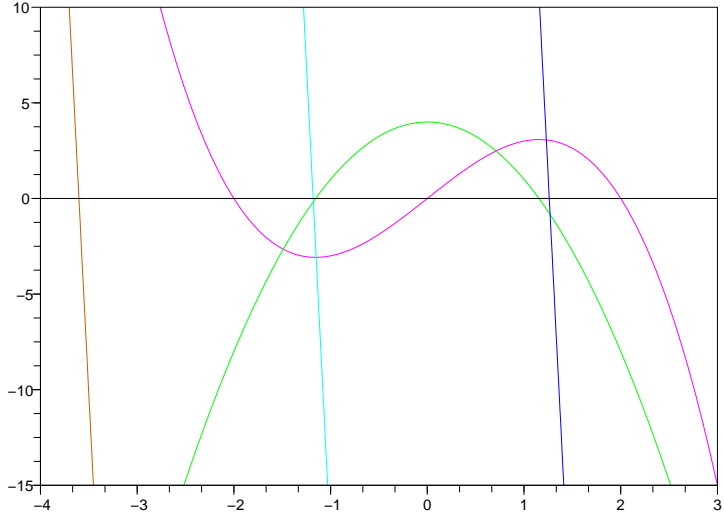}
\caption{Nullclines of the fast system}\label{nullcline}}
{\it \small The pink solide line corresponds to the $\dot{x}=0$ nullcline and the cyan solid line to the $\dot{y}=0$ nullcline for $X=0$ (no coupling). The green solid line corresponds to the curve $f'(x)=0$. The ($x_{0}^{-}, x_{0}^{+}$) roots of $f'(x)=-a_1\delta\!\approx\!0$ roughly correspond to the intersection points of the green line with the black horizontal line. As the value of $X$ varies, the cyan nullcline sweeps the x-axis within or outside the $[x_{0}^{-},x_{0}^{+}]$ interval, between extremal positions delimited by the brown and blue straight lines.}
\end{figure}

System \eref{eqdotx}-\eref{eqdotY} can be analyzed in the general setting of fast-slow
dynamics. The slow variable $X$ then enters the fast system \ref{eqdotx}-\ref{eqdoty} as a parameter. Bifurcations may thus arise as the $X$ parameter varies. The fast system (after changing time $t$ into ${\epsilon}t$) also exhibits fast-slow dynamics features due to the $\delta$ time scale:
\begin{subequations}
\begin{eqnarray}
\delta\dot{x}&=&-y+f(x)\label{dotxfast}\\
\dot{y}&=&a_{0}x+a_{1}y+a_{2}+cX \label{dotyfast}\\
\text{with }f(x)&=&\nu_{0}x^3+\nu_{1}x^2+\nu_{2}x
\end{eqnarray}
\end{subequations}
The fast nullcline $\dot{x}\!=\!0$ is a cubic function and the stationary
points are obtained as intersections of this cubic nullcline with
the other nullcline $\dot{y}\!=\!0$, a straight line which moves in the plane depending on the values of the slow variable $X$. Without loss of generality, we assume that $\lim_{x\to-\infty}f(x)=+\infty$. We also assume that the parameters $(a_{0},a_{1})$
are chosen in such a way that there is one single stationary point whatever the value of $X$ may be (see Figure \ref{nullcline}). Intersection points of the cubic nullcline with the line nullcline
are solutions of the equation:
\begin{equation}\label{roots}
f(x)+\frac{a_{0}}{a_{1}}x+\frac{\left(cX+a_2\right)}{a_1}=0.
\end{equation}

\noindent Let $x_{0}=s(X)$ denote the unique solution to equation \eref{roots}. We now derive from classical arguments the nature of the stationary point $(s(X),
f(s(X)))$. The eigenvalues of the linearized system are solutions of the characteristic polynomial:
$$\lambda^{2}-[\frac{f'(x_{0})+a_1}{\delta}]\lambda+
\frac{1}{\delta}[a_{0}+a_{1}f'(x_{0})]=0.$$
If the following conditions are fulfilled, two roots are complex conjugated with a negative real part.
$$
\left\{
\begin{array}{l}
\ds{\frac{1}{\delta}f'(x_{0})}+a_{1}<0\\
\ds{\frac{a_{0}}{\delta}}+\ds{\frac{a_{1}}{\delta}}f'(x_{0})>0\\
\left[\displaystyle{\frac{f'(x_{0})}{\delta}}-a_{1}\right]^{2}-4\displaystyle{\frac{a_{0}}{\delta}}<0,
\end{array}
\right.
$$

\noindent Let $x_{0}^{-}$  and $x_{0}^{+}$ denote the two roots of the equation:
$$\frac{1}{\delta}f'(x)+a_{1}=0.$$
Then, if $s(X)<x_{0}^{-}$, the stationary point is a stable focus. If
$X$ varies in such a way that $x_{0}=s(X)$ crosses the value
$x_{0}^{-}$, the system \eref{dotxfast}-\eref{dotyfast} undergoes a Hopf bifurcation, and the
stationary point becomes unstable as a stable limit cycle appears. This stable limit cycle disappears when $x_{0}>x_{0}^{+}$
and the stationary point becomes stable again. This reasoning is illustrated on Figure \ref{nullcline}.
\subsubsection{Limit cycle of the slow system}
The dynamics of the $X$ variable is such that, if the straight-line nullcline sweeps the (x,y) phase plane (from left to right and right to left) periodically, the Hopf bifurcations occur themselves periodically.\\

\begin{figure}[H]
{\centering
\includegraphics[scale=0.6]{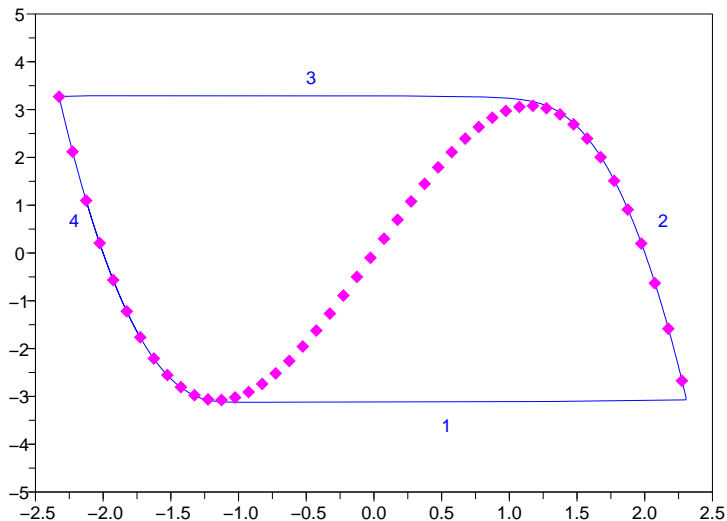}
\caption{Slow $(X,Y)$ limit cycle}\label{limXY}}
{\it\small The blue line corresponds to the limit cycle in the $X,Y$ plane. The slow parts of the cycle correspond to branches of the cubic nullcline represented by pink diamonds (phases 2 and 4), the fast parts to the jumps from one branch of the cubic to the other one (phases 1 and 3). }
\end{figure}

The qualitative analysis of the slow dynamics is analogous to the fast one. We assume that the parameters $(b_{0},b_{1},b_{2})$
are fixed so that there is one single stationary point $(X_{0}, F(X_{0}))$, which is an unstable focus (allowing the slow system to displays a stable limit cycle):

$$
\left\{
\begin{array}{l}
\ds{\frac{1}{\gamma}F'(X_{0})}+b_{1}>0\\
\ds{\frac{b_{0}}{\gamma}}+\ds{\frac{b_{1}}{\gamma}}F'(X_{0})>0\\
\left[\displaystyle{\frac{F'(X_{0})}{\gamma}}-b_{1}\right]^{2}-4\displaystyle{\frac{b_{0}}{\gamma}}<0,
\end{array}
\right.
$$
$\text{where }F(x)=\mu_{0}X^3+\mu_{1}X^2+\mu_{2}X$.
\noindent The oscillation associated to this limit cycle is of relaxation
type (as in van der Pol system). The dynamics along the limit
cycle displays slow and fast parts alternatively (see Figure \ref{limXY}).

\subsubsection{Dynamics of the coupled system}\label{coupled}
The different phases of the limit cycle exhibited by the slow regulatory system (see Figure \ref{limXY}) drives the global behavior of the fast GnRH system (see Figure \ref{outputfhn}):
\begin{enumerate}
\item the first phase of the cycle, where $X$ increases abruptly, corresponds to the ascending part of the surge;
\item the second phase of the cycle, where $X$ decreases slowly, corresponds to the duration of the surge;
\item the third phase of the cycle,  where $X$ decreases abruptly, corresponds to the decreasing part of the surge;
\item the fourth and most lasting phase of the cycle,  where $X$ increases slowly, encompasses two different GnRH secretion patterns:
\begin{enumerate}
\item as long as $X\leq x_{0}^{+}$, GnRH level is almost constant, since the fast system admits a stable steady state and the sweeping dynamics of the straight line $\dot{y}=0$ lies in a slow phase. Hence this phase explains the existence of a plateau after the surge. 
\item when $X> x_{0-}^{+}$ the pulsatility of GnRH is recovered and the pulse frequency increases with $X$.
\end{enumerate}
\end{enumerate}

\section{Numerical simulations}\label{simul}

The numerical values of the model parameters can be constrained by physiological specifications regarding the features (frequency, amplitude, plateau length) of the GnRH secretory patterns \cite{evans_02}. The GnRH output should be characterized by:
\begin{itemize}
\item the plateau-length to frequency ratio for the fast oscillations; 
\item the pulse-amplitude to surge-amplitude ratio; 
\item the surge-frequency to pulse-frequency ratio; 
\item the surge-duration to whole-ovarian-cycle duration.
\end{itemize}

\begin{figure}[H]
{\centering
\includegraphics[scale=0.6]{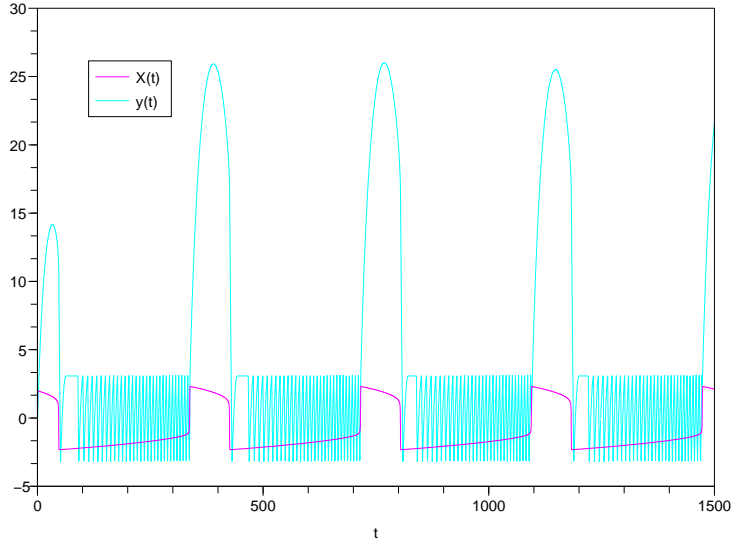}
\caption{Outputs from the coupled systems}\label{outputfhn}}
{\it\small \begin{center} pink line: $X(t)$, cyan line:  $y(t)$\end{center}}
\end{figure}

\begin{figure}[H]
\begin{minipage}{0.65\linewidth}
\includegraphics[scale=0.4]{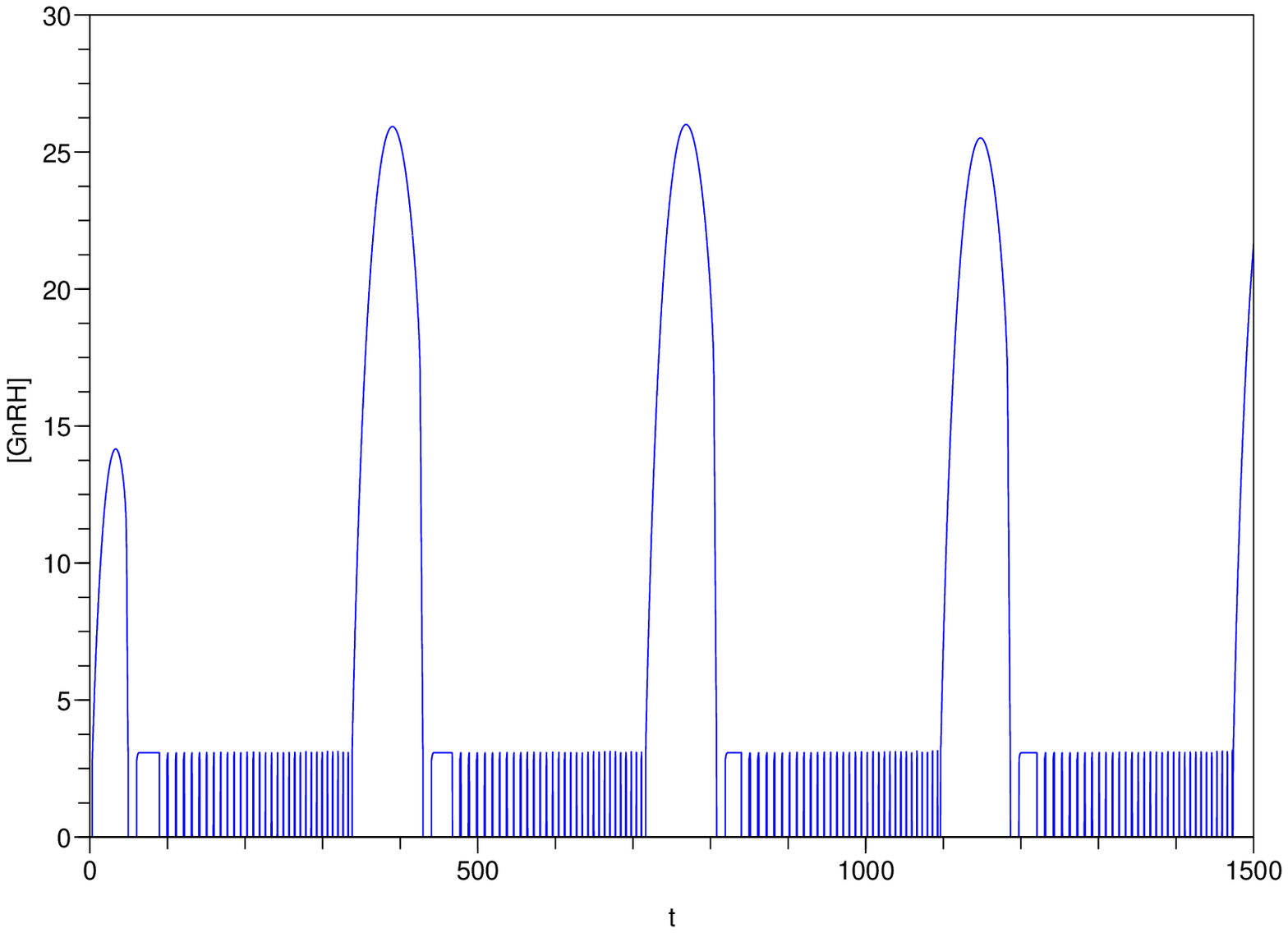}
\caption{GnRH secretion pattern}\label{outputgnrh}
{\small\it The secretory activity of GnRH neurons occurs for threshold ionic activity $z(t)=\chi_{\left\{y(t)>y_s\right\}}$}
\end{minipage}
\begin{minipage}{0.02\linewidth}
\end{minipage}
\begin{minipage}{0.32\linewidth} 
\includegraphics[scale=0.2]{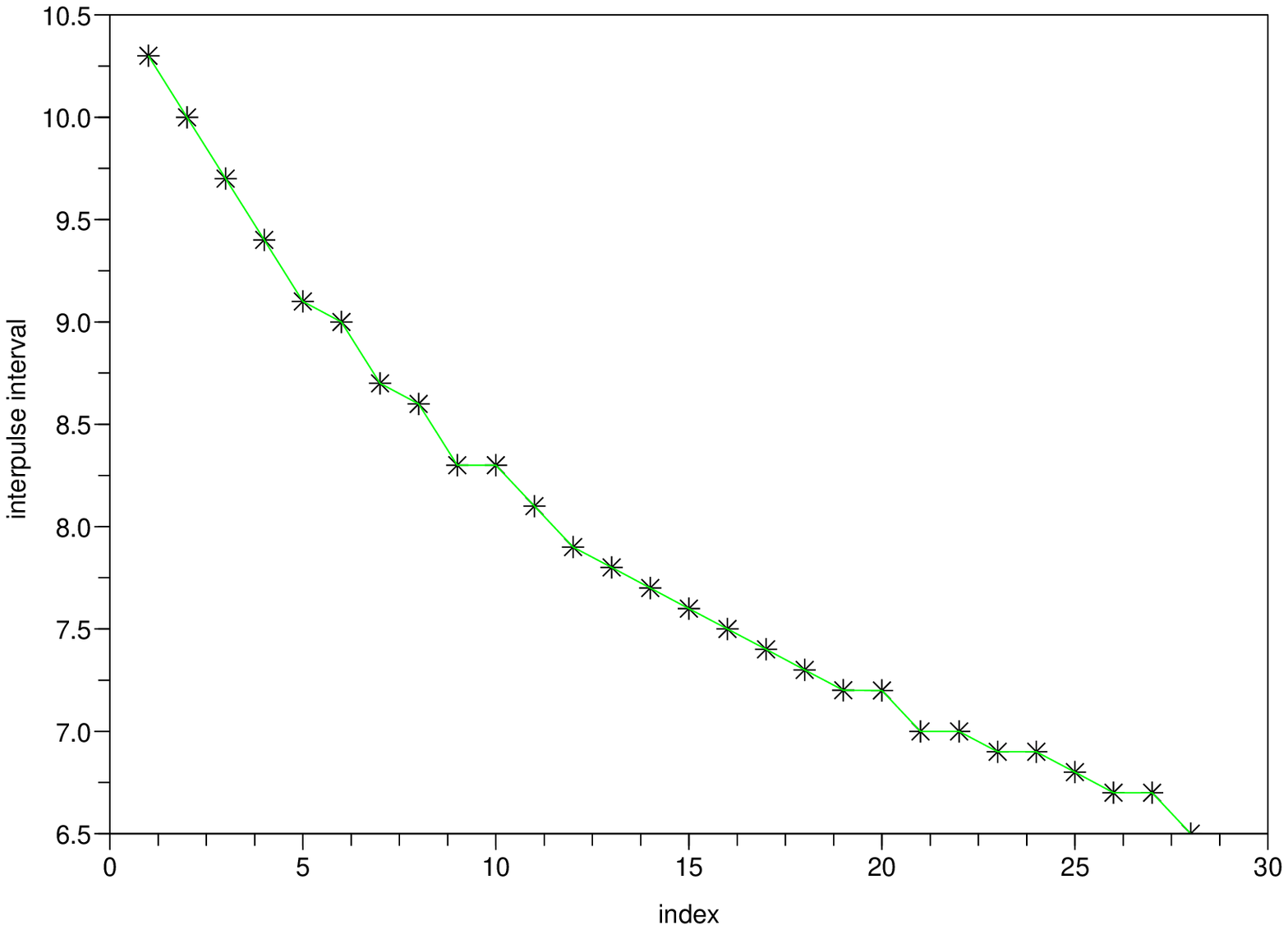}

\includegraphics[scale=0.2]{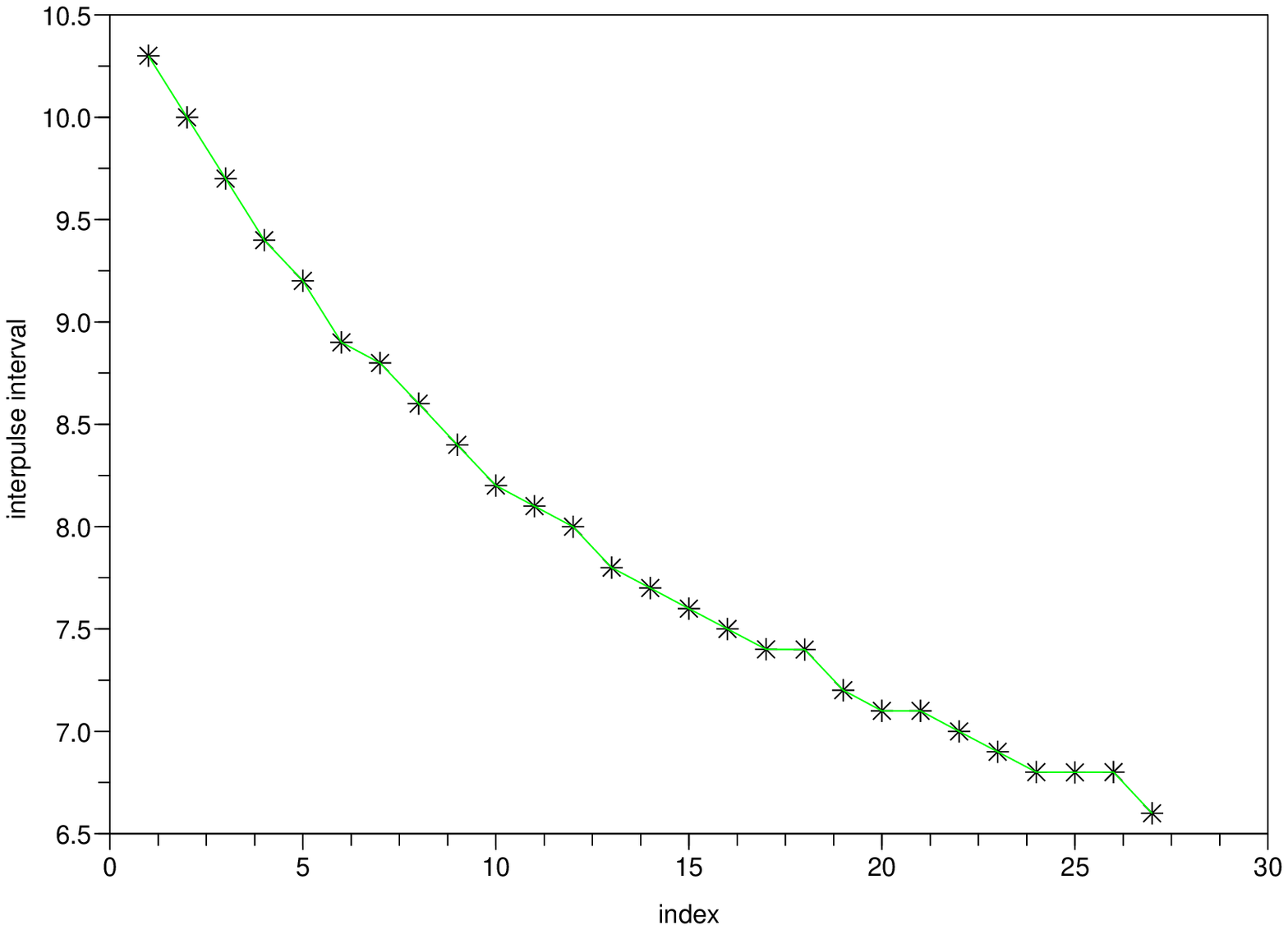}
\caption{Decrease in interpulse intervals from the luteal to the follicular phase (top: from time 100 to time 337, bottom: from time 475 to time 710)}\label{periodes}
\end{minipage}
\end{figure}

\noindent A set of parameters subject to such constraints is displayed on Table \ref{table1}, and the corresponding model outputs are illustrated on Figures \ref{outputfhn} and \ref{outputgnrh}.

\begin{table}[H]
\centering
\begin{tabular}{c c| c c| c c}
$\varepsilon$&1/40&
$\delta$&1/80&
$\gamma$&1\\
$a_0$&1&
$a_1$&0.01&
$a_2$&1.18\\
$b_0$&1&
$b_1$&0.04&
$b_2$&1\\
$\nu_0$&-1&
$\nu_1$&0&
$\nu_2$&4\\
$\mu_0$&-1&
$\mu_1$&0&
$\mu_2$&4\\
$c$&1.05&$y_s$&2.8&
\end{tabular}

\caption{Numerical values of the model parameters}\label{table1}
\end{table} 

The main qualitative features captured by the model consist in:
\begin{enumerate} 
\item the cyclic transition from a pulse to a surge secretion pattern, which occurs after a short transitory period and seems not to be subject to initial conditions;
\item a delay before resumption of pulsatility; 
\item the increase in pulse frequency from the luteal (post-surge) phase to the follicular (pre-surge) phase (see Figure \ref{periodes}).
\end{enumerate}

\noindent If the two first properties (pulse to surge alternating and pulsatility resumption delay) were expected from the study derived in section \ref{modeldesign}, the third one (frequency increase) remains unexplained at this point, even if it is consistent with endocrinological data. Besides, the global behavior of the system is not strictly periodic, since both the duration of the delay and GnRH surge amplitude are not constant.  The non-strict periodicity is a nice feature for a deterministic model in a biological context and it also reserves further analysis. We now explain the mechanisms underlying those properties from the bifurcation analysis of a 2-scaled system derived from the model.

\section{Bifurcation analysis of a model-derived 2-scaled system} \label{bifurcation}

Our approach is adapted from the "geometrical dissection" \cite{borisyuk}, that has been successfully applied to several models in Computational Neurosciences, especially those dealing with bursting oscillations.

\noindent In classical slow/fast systems, the slow variable is "frozen" and appears as a parameter in the study of the bifurcations of the fast system. In a similar way, we consider the fast 3 dimensional system with 2 time scales:

\begin{subequations}
\begin{eqnarray}
\delta\dot{x}&=&-y+f(x)\label{dotxfast2}\\
\dot{y}&=&a_{0}x+a_{1}y+a_{2}+cX \label{dotyfast2}\\
\gamma\dot{X}&=&-Y+F(X)\label{dotXfast2}
\end{eqnarray}
\end{subequations}

\noindent where $Y$ acts as a varying parameter. Below, we restrict our study to the case where the other parameter values are fixed and close to those of Table \ref{table1}. This fast system breaks into an independent, 1D system (\ref{dotXfast2}), and a  2D system (\ref{dotxfast2}-\ref{dotyfast2}) forced by the 1D system. \\

\noindent Depending on $Y$ value, system (\ref{dotXfast2}) displays either one of the 2 possible attracting points (denoted respectively by $X_{-}(Y)$ and $X_{+}(Y)$), or these two attractors separated by a repulsive point (denoted by $X_{0}(Y)$). Saddle-node bifurcations occur when $Y$ value equals one of the local extrema of the cubic function: $(X,Y)=\pm(2/\sqrt{3},16/(3\sqrt{3}))=\pm(1.15,3.08)$. The whole bifurcation analysis of the 3D system is summarized in Table \ref{tablebif}.\\

\begin{center}
\begin{table}[H]
{\scriptsize
\begin{tabular}{|c | c  c  c | c  c  c |c  c  c |c  c  c |}
\hline
&&&-3.08&&&0&&&3.08&&&$+\infty$\\
Y&&$\nearrow$&&&$\nearrow$&&&$\nearrow$&&&$\nearrow$&\\\
&$-\infty$&&&-3.08&&&0&&&3.08&&\\
\hline
&$+\infty$&&&$$&&&&&&///&///&///\\
$X_{+}(Y)$&&$\searrow$&&&$\searrow$&&&$\searrow$&&///&///&///\\
&&&&&&&&&&///&///&///\\
\hline
&&&&&&&&&&///&///&///\\
$x_{+}(Y)$&&$\nearrow$&&&$\nearrow$&&&$\nearrow$&&///&///&///\\
&$-\infty$&&&&&&&&&///&///&///\\
\hline
&\multicolumn{7}{r}{attractive stationary point}&&&\!\!\!\!\!\!\!\!\!{\bf (1)}$\;$///&///&///\\
\hline
&///&///&///&&&&&&&///&///&///\\
$X_{0}(Y)$&///&///&///&&$\nearrow$&&&$\nearrow$&&///&///&///\\
&///&///&///&&&&&&&///&///&///\\
\hline
&///&///&///&&&&&&&///&///&///\\
$x_{0}(Y)$&///&///&///&&$\searrow$&&&$\searrow$&&///&///&///\\
&///&///&///&&&&&&&///&///&///\\
\hline
&///&///&///&\!\!\!\!\!\!\!\!\!\!\!\!\!\!\!\!{\bf (2) } &\!\!\!\!\!\!\!\!repulsive stationary point&&\!\!\!\!\!\!\!\!\!\!\!\!\!\!\!\!\!\!\!\!\!\!\!\!\!\!\!\!\!\!\!\!\!\!\!{\bf (4) } &\!\!\!\!\!\!\!\!\!\!\!\!\!\!\!\!saddle point&&\!\!\!\!\!\!\!\!{\bf (5)} ///&///&///\\
&///&///&///&\!\!\!\!\!\!\!\!\!\!\!\!\!\!\!\!\!\!{\bf (3) }& \!\!\!\!\!\!\!\!hyperbolic periodic orbit&&\!\!\!\!\!\!\!\!\!\!{\bf} &&&///&///&///\\
\hline
&///&///&///&&&&&&&&&\\
$X_{-}(Y)$&///&///&///&&$\searrow$&&&$\searrow$&&&$\searrow$&\\
&///&///&///&&&&&&&&&\\
\hline
&///&///&///&&&&&&&&&\\
$x_{-}(Y)$&///&///&///&&$\nearrow$&&&$\nearrow$&&&$\nearrow$&\\
&///&///&///&&&&&&&&&\\
\hline
&///&///&///&\multicolumn{4}{l}{\!\!\!\!\!\!\!\!{\bf (6)} $\qquad\qquad$ $\qquad\qquad\qquad$attractive periodic orbit}&&&\!\!\!\!\!\!\!\!\!\!\!\!\!\!\!\!\!\!\!{\bf (8)}&\!\!\!\!\!\!\!\!\!\!\!\!\!\!attractive&\\
&///&///&///&\multicolumn{4}{l}{\!\!\!\!\!\!\!\!{\bf (7)} $\qquad\qquad$ $\qquad\qquad$ $\qquad\qquad\qquad$saddle point}&&&\!\!\!\!\!\!\!\!\!\!\!\!\!\!\!\!\!\!\!{\bf}&\!\!\!\!\!\!\!\!\!\!\!\!\!\!stationary&\!\!\!\!\!\!\!\!\!\!point\\
\hline
\end{tabular}
\caption{Bifurcations of the 3D system according to $Y$ value}\label{tablebif}
{\it\small The numbers superimposed on the vertical straight lines separating the columns correspond to bifurcations of the following type\\(1)-(3)-(5)-(7): saddle-node, (2)-(6): saddle-node of periodics, (4)-(8): Hopf}}
\end{table}
\end{center}

\noindent We now go back to the 4D system (\ref{eqdotx})-(\ref{eqdotY}) to explain the sequence of phases listed in subsection \ref{coupled}.
\begin{itemize}
\item In phase 1, $Y$ decreases and system (\ref{dotxfast2})-(\ref{dotXfast2}) displays the single attractive node associated to $X_{+}(Y)$, corresponding to the ascending part of the surge;
\item As $X_{+}(Y)$ decreases slowly, the solution of the 4D system remains close to the attractive node, corresponding to the duration of the surge (phase 2), until this node disappears through a saddle-node bifurcation. Then the solution switches to the other attractive node $X_{-}(Y)$, corresponding to the decreasing part of the surge (phase 3);
\item In phase 4, as $X_{-}(Y)$ increases slowly, the solution remains close to the attractive node associated to $X_{-}(Y)$, corresponding to the plateau. Eventually, this attractive node disappears into an attractive periodic orbit initiating the pulsatile phase;
\item As phase 1 starts again, $X$ speeds up and the pulse frequency increases. At some point the attractive periodic orbit disappears into a saddle node of periodics. The solution of the 4D system then jumps back to the single attractive node associated to $X_{+}(Y)$ and recovers the ascending part of the surge.
\end{itemize}

\noindent Hence this bifurcation analysis has shown that the alternating pulse and surge pattern is based fundamentally on an underlying hysteresis loop. 

\section{Identification of parameter targets for steroid control}\label{control}
In this section we derive some tools that will be useful in controlling  the amplitude and frequency of oscillations in either the slow regulatory system or the GnRH system, hence to mimick either indirect or direct effects of steroid feedback. 
\subsection{Amplitude and frequency of the slow regulatory oscillations}
\subsubsection{Amplitude of the slow limit cycle}
Let us first derive a parametric expression of the $F(X)$ cubic as a function of its extrema. Denote by $-\alpha$ and $\beta$ the roots of $F'(X)=0$. Without loss of generality, we assume $\alpha>0$ and $\beta>0$ and choose the cubic function $F(X)$ so that it passes through the origin and takes minus unity as coefficient for the higher (cubic) power. 
\begin{eqnarray*}
F(X)&=& -X^3+\mu_{1}X^2+\mu_{2}X\quad\text{ with }\mu_1>0,\, \mu_2>0 \text{ and } \mu_0=-1\\
F'(X)&=&-3X^2+2\mu _{1}X+\mu_{2}\\
&=&-3(X+\alpha)(X-\beta)\\
&=&-3\left[X^2+X(\alpha-\beta)-\alpha\beta\right]\\
F(X)&=&-X^3-\frac{3}{2}(\alpha-\beta)X^2+3\alpha\beta X
\end{eqnarray*}

\begin{figure}[H]
{\centering
\includegraphics[scale=0.65]{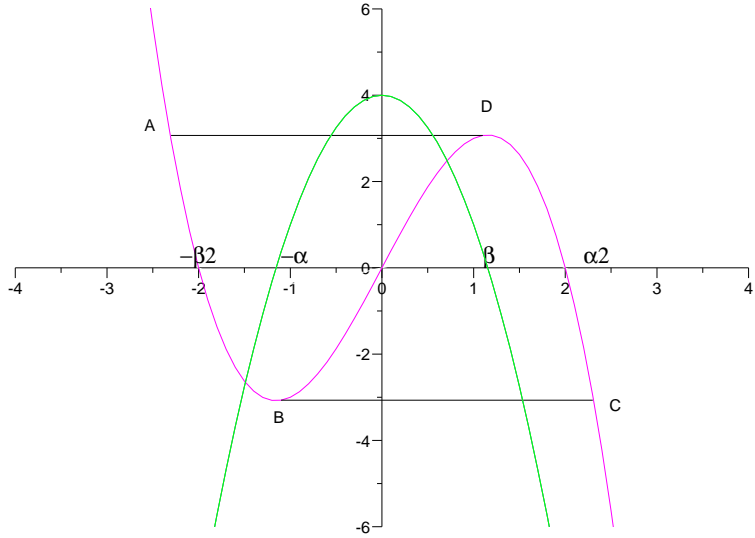}
\caption{Amplitude and period of the slow limit cycle $(X,Y)$}\label{cubic_alpha_beta}}
{\it \small The period of the slow $(X,Y)$ limit cycle is computed from the time taken to go along the path {\bf A}$(-\beta_2, F(-\beta_2))$~- {\bf B}$(-\alpha, F(-\alpha))$~- {\bf C}$(\alpha_2, F(\alpha_2))$~- {\bf D} $(\beta, F(\beta))$, with $F'(\alpha)\!=\!F'(\beta)\!=\!0$, $F(\alpha_2)\!=\!F(\alpha)$, $F(\beta_2)\!=\!F(\beta)$. The global amplitude of $X$ corresponds to $\alpha_2+\beta_2$, the surge-related amplitude to $\alpha_2+\alpha\!=\!\beta+\beta_2$}
\end{figure}

\noindent Hence $\alpha$ and $\beta$ are the positive roots of
$$\begin{cases}
3\alpha\beta=\mu_2&\\
3\beta^2-2\mu_1\beta-\mu_2=0&
\end{cases}$$

$$\text{so that } \qquad\qquad\qquad\qquad\alpha=\frac{\mu_2}{\mu_1+\sqrt{\mu_1^2+3\mu_2}}\qquad \text{ and }\qquad\beta=\frac{\mu_1+\sqrt{\mu_1^2+3\mu_2}}{3}$$
To compute the amplitude of the $X$ variable, we now seek for the points of the cubic verifying either $F(X)\!\!=\!\!F(-\alpha)$ or 
 $F(X)\!\!=\!\!F(\beta)$, and $F'(X)\!\!\neq \!\!0$ (see Figure \ref{cubic_alpha_beta}). 
$$F(X)=F(-\alpha)=-\frac{1}{2}\alpha^2(\alpha+3\beta) \Leftrightarrow (X-\alpha)^2\left[X-\frac{1}{2}(\alpha+3\beta)\right]=0$$ 
$$F(X)=F(\beta)=\frac{1}{2}\beta^2(\beta+3\alpha) \Leftrightarrow (X-\beta)^2\left[X+\frac{1}{2}(\beta+3\alpha)\right]=0$$
Define 
$$ \alpha_2\equiv\frac{1}{2}(\alpha+3\beta)\;\text{ and }\;\beta_2\equiv\frac{1}{2}(\beta+3\alpha)$$
Then $F(\alpha_2)=F(-\alpha)$, $F(-\beta_2)=F(\beta)$, and the global amplitude of $X$ is thus given by $\alpha_2+\beta_2={2}(\alpha+\beta)$, while the fast jump amplitude (related to the surge amplitude in the GnRH system) is given by $\alpha_2+\alpha=\beta+\beta_2=\frac{3}{2}(\alpha+\beta)$. In the symetric case where $\alpha=\beta$, those amplitudes respectively reduce to $4\alpha$ and $3\alpha$ . 

\subsubsection{Period of the slow limit cycle}
The frequency of the regulatory oscillations drives the frequency of the GnRH surge. The period of the slow cycle (i.e the time taken to move along the whole cycle) in $X$ indeed corresponds to the period of the GnRH surge. 

\noindent Let us denote by $T$ the period of the $(X,Y)$ limit cycle.
The time $T_X$ taken to move along the slow part (ie along the branches of the cubic function $F$) of the cycle can be computed as:

\begin{eqnarray*}
T_X&=&\int_{F(-\beta_2)}^{F(-\alpha)}{\displaystyle{\frac{dY}{b_0 X+b_1 Y+b_2}}}+\int_{F(\alpha_2)}^{F(\beta)}{\displaystyle{\frac{dY}{b_0 X+b_1 Y+b_2}}}\\
&=&\int_{-\beta_2}^{-\alpha}{\displaystyle{\frac{F'(X)dX}{b_0 X+b_1 F(X)+b_2}}}+\int_{\alpha_2}^{\beta}{\displaystyle{\frac{F'(X)dX}{b_0 X+b_1 F(X)+b_2}}}\\
\end{eqnarray*}

\noindent It is worth noticing that, within the $T_X$ duration, the time taken to climb up the descending (right) branch of the cubic (integration from $F(\alpha_2)$ to $F(\beta)$) corresponds to the duration of the surge.

\noindent The time $T_Y$ taken to move along the quasi-horizontal fast parts (corresponding to the jumps from one branch of the cubic function to another) of the $(X,Y)$ limit cycle can be computed as
\begin{eqnarray*}T_Y&=&\int_{-\alpha}^{\alpha_2}{\displaystyle{\frac{\gamma dX}{F(X)-Y}}}+\int_{\beta}^{-\beta_2}{\displaystyle{\frac{\gamma dX}{F(X)-Y}}}\\
\end{eqnarray*}
\noindent Alternatively, $T_Y$ can be estimated as $\approx 2\epsilon\gamma\,T_X$, according to the ratio of time constants in equations (\ref{eqdotX}) and (\ref{eqdotY}).

\noindent The whole $(X,Y)$ limit cycle period can be finally computed from $T=T_X+T_Y\;.$

\subsubsection{Changes in amplitude and frequency according to a prescribed ratio}
From the previous analysis, we are able to assess numerically the frequency and amplitude of the slow $(X,Y)$ limit cycle for any arbitrary parameter set. We now make a further remark. 
\noindent Given the system: 
\begin{eqnarray*}
\gamma\dot{X}&=&F(X)-Y\\
\dot{y}&=&b_0 X+b_1 Y+b_2,
\end{eqnarray*}

\noindent one can one find $\left\{\bar{\gamma},\bar{f},\bar{b_0},\bar{b_1},\bar{b_2}\right\}$ to transform it into another system, 
\begin{subequations}
\begin{eqnarray}
\bar{\gamma}\dot{X}&=&\bar{F}(X)-Y\label{fhn1}\\
\dot{Y}&=&\bar{b_0}X+\bar{b_1}Y+\bar{b_2}\label{fhn2}
\end{eqnarray}
\end{subequations}
\noindent oscillating with prescribed $\lambda_1$ frequency and $\lambda_2$ amplitude ratios.\\

This change of variables leads to
\begin{eqnarray*}
\bar{\gamma}&=&\lambda_1 \gamma\\
\bar{b_0}&=&\lambda_1 b_0\\
\bar{b_1}&=&\lambda_1 b_1\\
\bar{b_2}&=&b_2\frac{\lambda_1}{\lambda_2}\\
\bar{\mu_0}&=&\lambda_2^2\mu_0\\
\bar{\mu_1}&=&\lambda_2\mu_1
\end{eqnarray*}
In other words, the new system reads
\begin{eqnarray*}
\gamma\lambda_1\dot{X}&=&-Y+\lambda_2^2 \mu_0 x^3 +\lambda_2\mu_1 x^2+\mu_2 x\\
\dot{Y}&=&\lambda_1\left(b_0X+b_1 Y+\frac{b_2}{\lambda_2}\right)
\end{eqnarray*}

\noindent The transformation does not modify the number of intersection points between the $\dot{x}=0$ and $\dot{y}=0$ nullclines, so that the qualitative behavior of the system is preserved.

\subsection{Targeting parameters for steroid effect}
The model allows to distinguish between two possible pathways for steroid feedback on GNRH secretion, either directly, by acting on the parameters of the faster system, or indirectly, by acting on those of the slower one. The former way is dedicated to the control of the frequency and amplitude of GnRH pulses, while the latter is dedicated to the control of the onset time and size of the GnRH surge. 

\subsubsection{Steroid-like direct effects}
\begin{figure}[H]
\begin{minipage}{0.65\linewidth}
\includegraphics[scale=0.4]{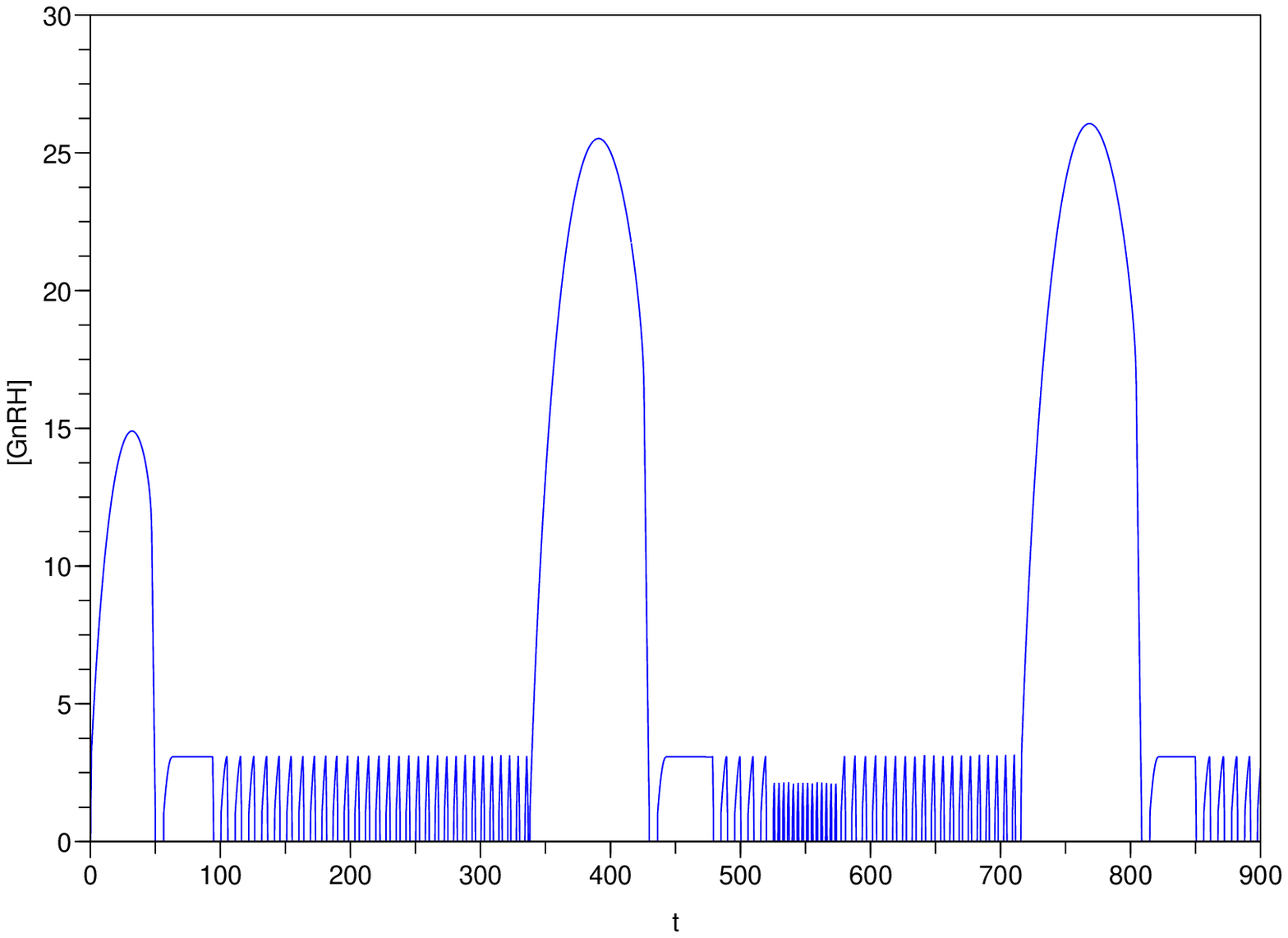}
\end{minipage}
\begin{minipage}{0.01\linewidth}
\end{minipage}
\begin{minipage}{0.33\linewidth}
\includegraphics[scale=0.2]{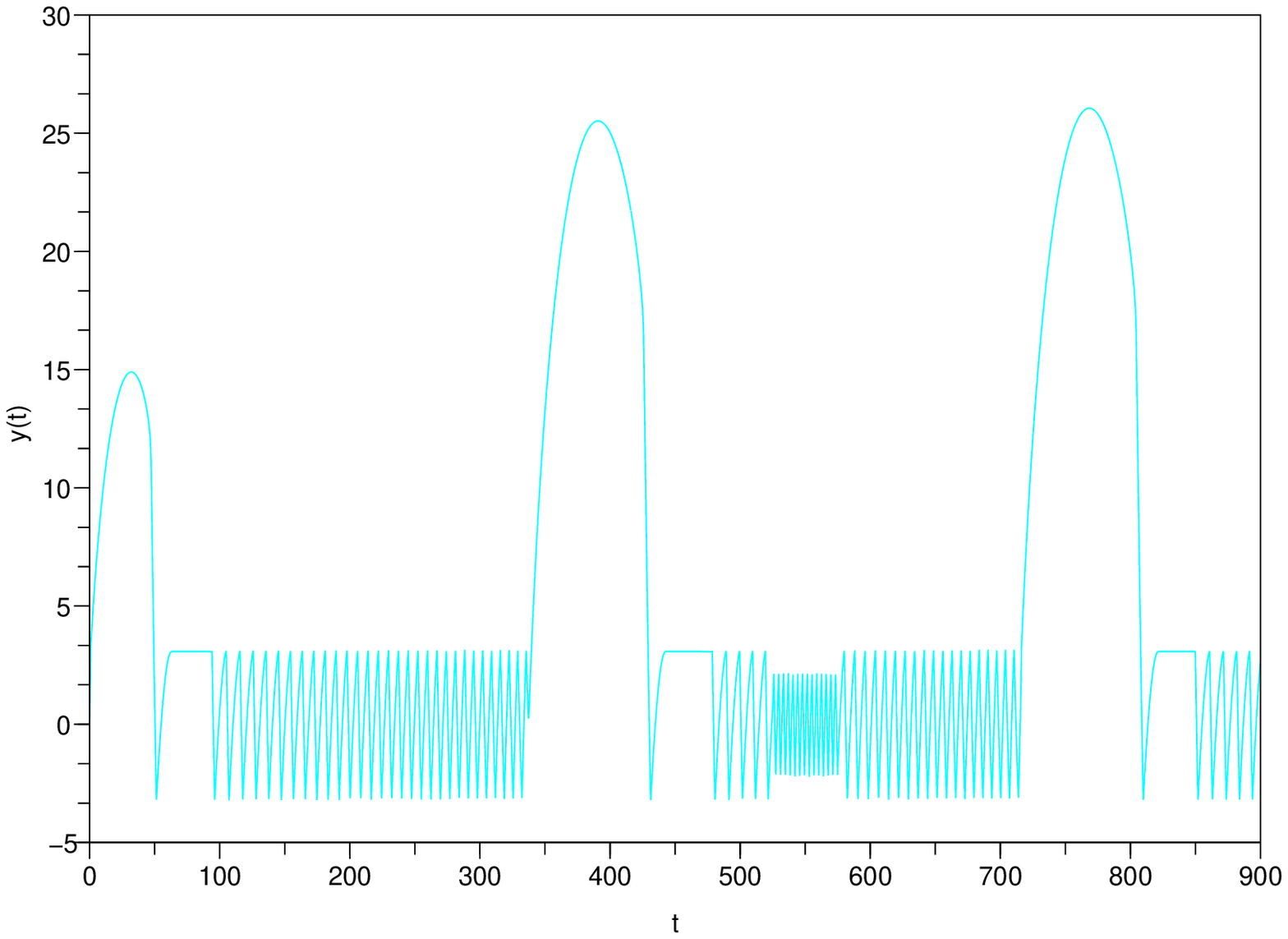}

\includegraphics[scale=0.2]{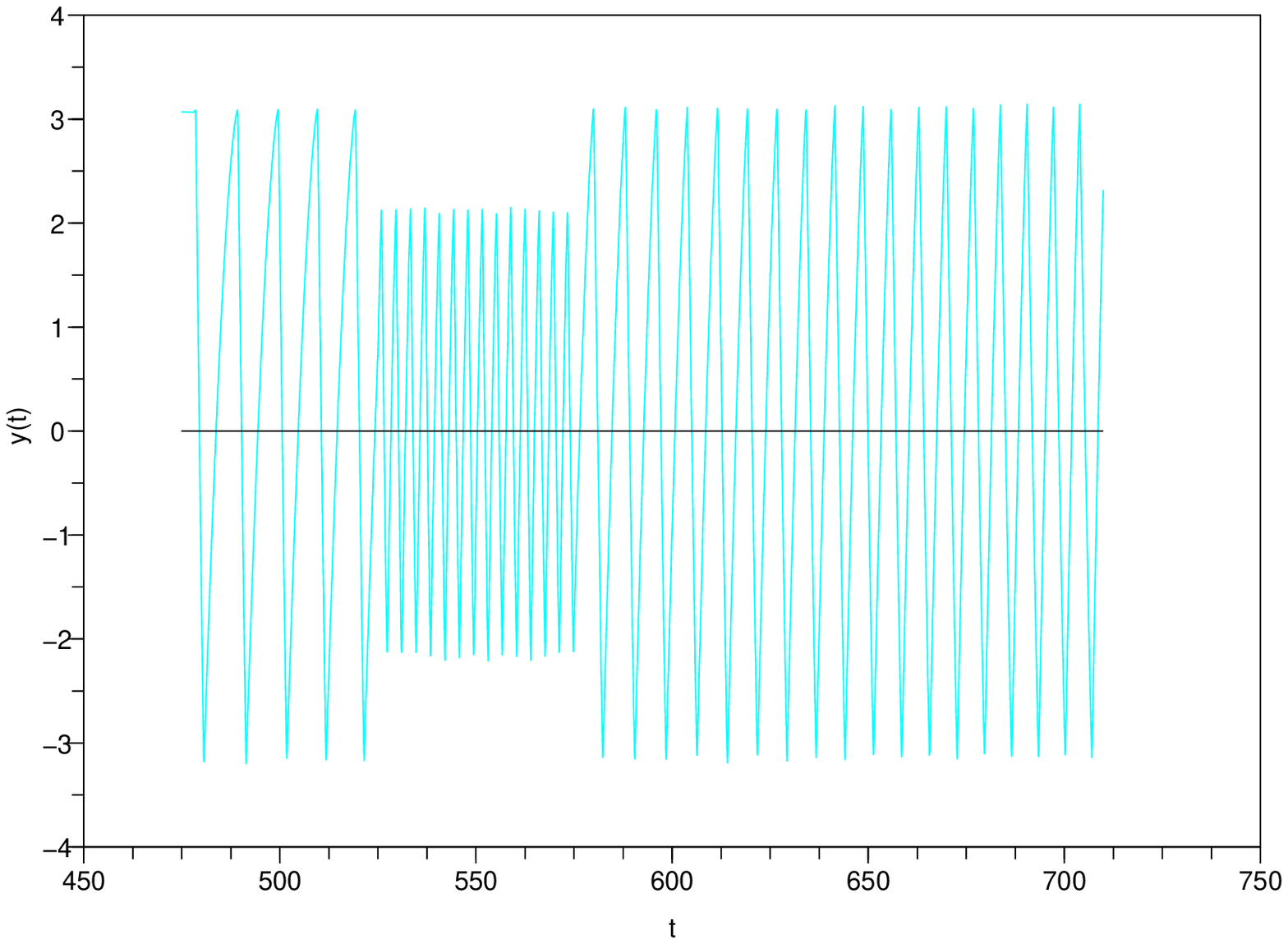}
\end{minipage}
\caption{Direct, estradiol-like effect, on portal GnRH output (left) and $y$ output (top right and zoomed on bottom right). The parameter values were acutely altered in a bolus way from time 525 to 575}\label{effetdirectE}
\end{figure}

Targeting adequate model parameters in the fast system in an acute way allows to alter transiently the amplitude and frequency of GnRH pulses. \\
Figures \ref{effetdirectE} illustrates the effect of a parameter-targeting bolus ($\lambda_1=2, \lambda_2=1.5$), from time 525 to 575 leading to a frequency-increased, amplitude-decreased pulsatile regime. Such an output can be compared with experimental results gathered in \cite{NPEvans1994}\footnote{where estradiol effects on GnRH secretory characteristics are summarized in Figure 3 and GnRH portal time series are displaid on Figure 2}. These effects consist in a reduction of GnRH pulse amplitude and a stimulatory action on GnRH pulse frequency, as accounted by the model results. 

\begin{figure}[H]
\begin{minipage}{0.65\linewidth}
\includegraphics[scale=0.4]{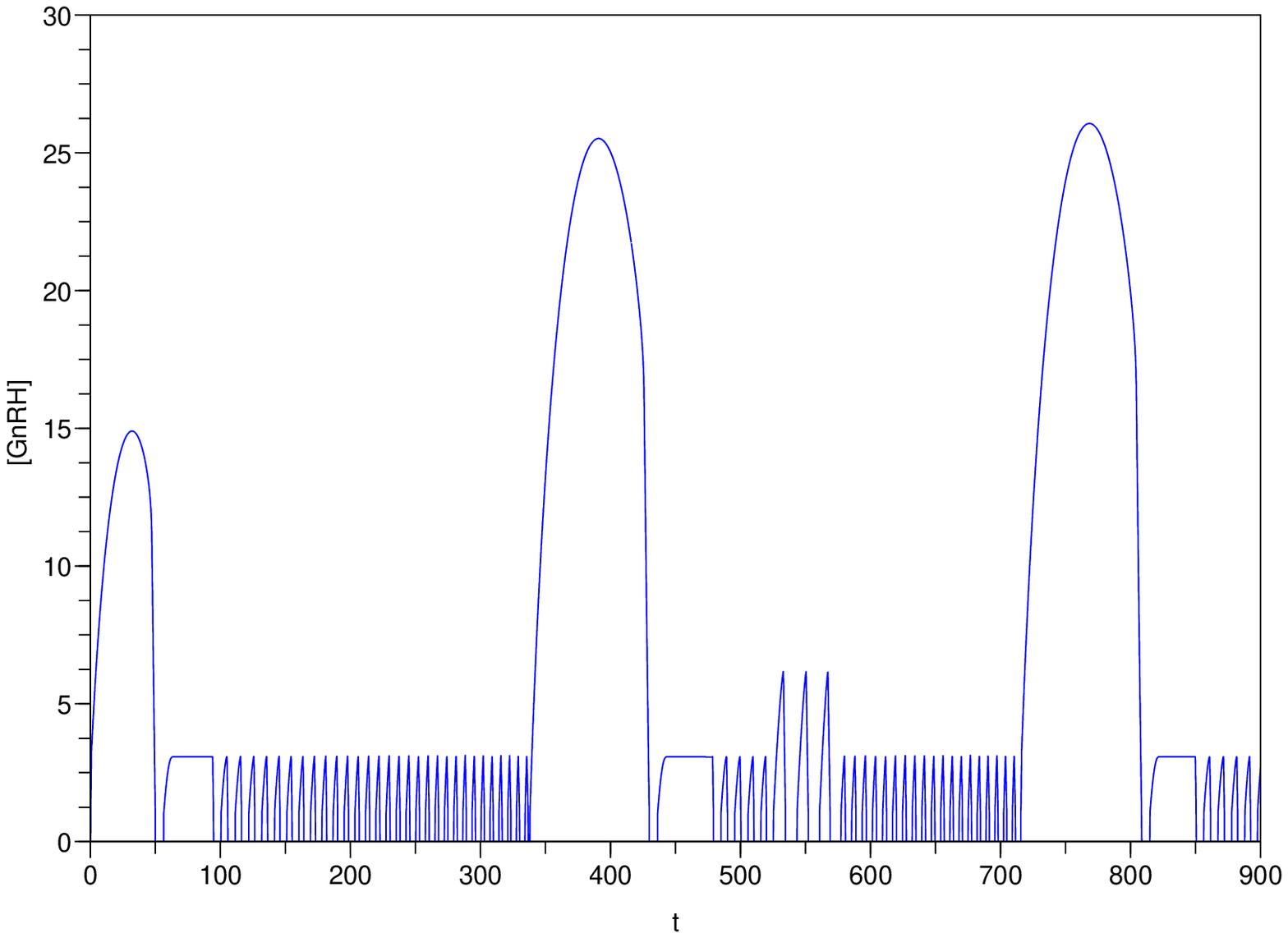}
\end{minipage}
\begin{minipage}{0.01\linewidth}
\end{minipage}
\begin{minipage}{0.33\linewidth}
\includegraphics[scale=0.2]{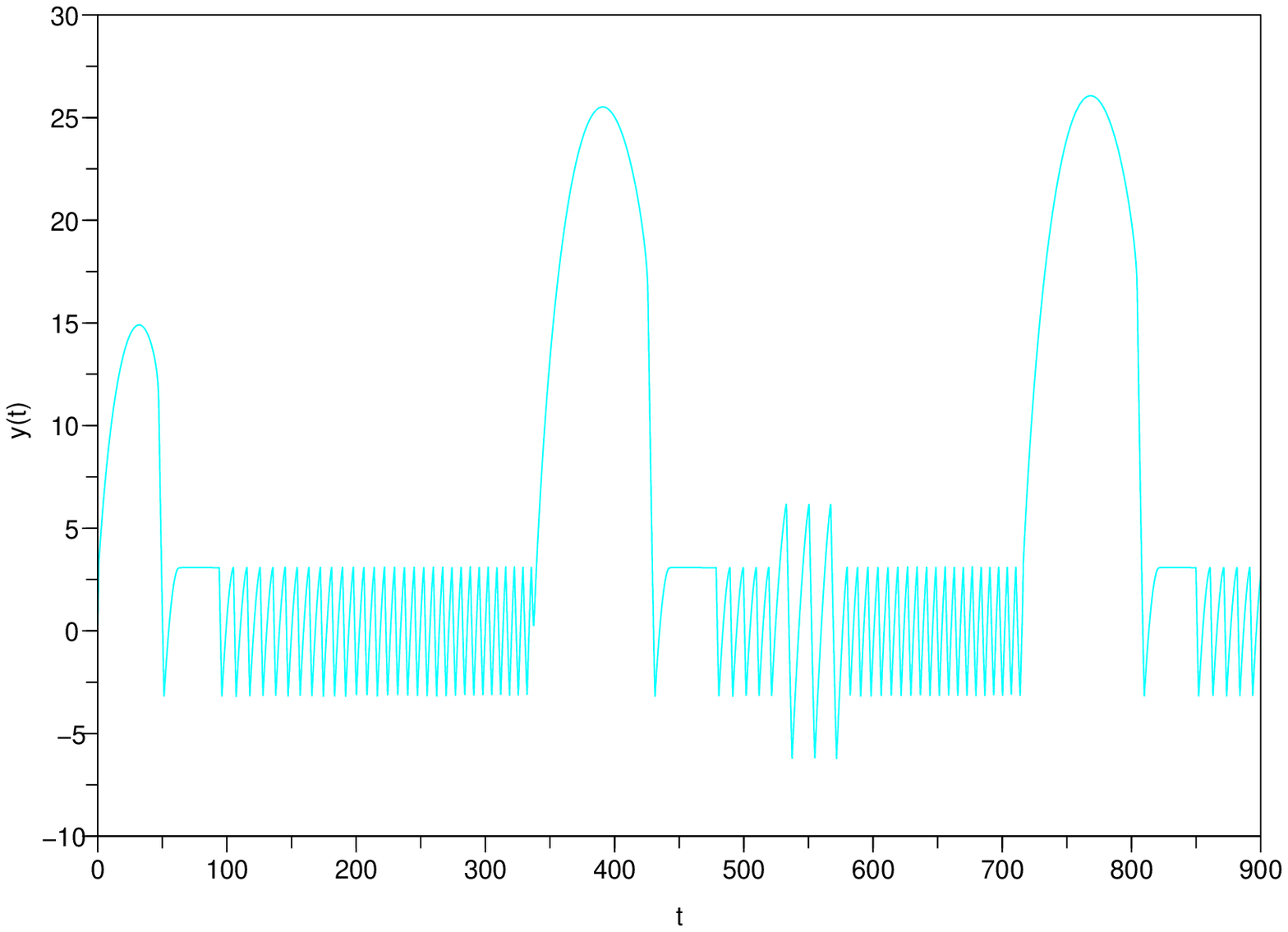}

\includegraphics[scale=0.2]{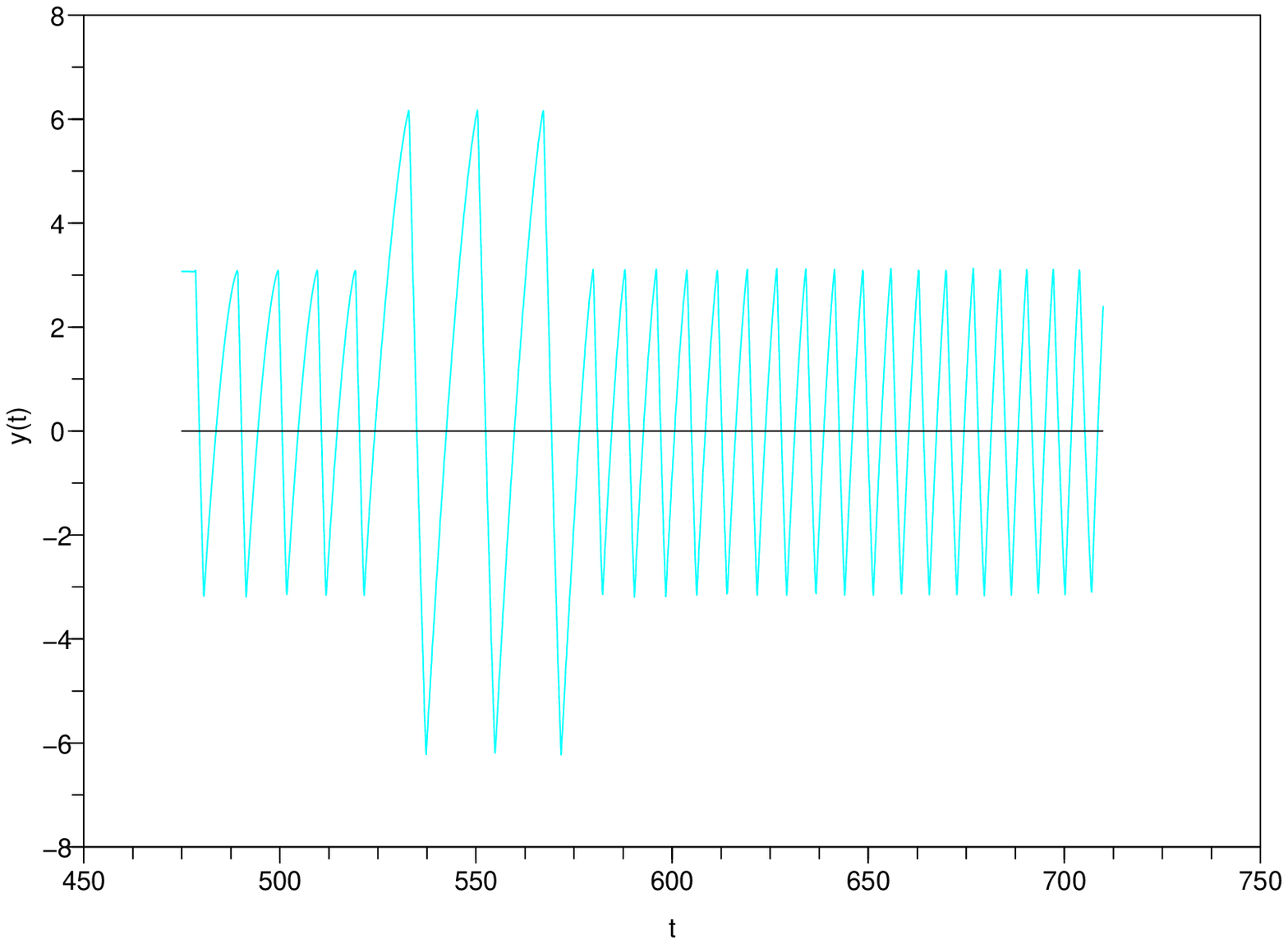}
\end{minipage}
\caption{Direct progesterone-like effect on portal GnRH ouput (left) and $y$ output (top right and zoomed on bottom right). The parameter values were acutely altered in a bolus way from time 525 to 575}\label{effetdirectP}
\end{figure}

Figures \ref{effetdirectP} illustrates the effect of a parameter-targeting bolus ($\lambda_1=0.5, \lambda_2=0.5$), from time 525 to 575 leading to a frequency-decreased, amplitude-increased pulsatile regime. Such an output can be compared by experimental results gathered in \cite{moenter_91}\footnote{where progesterone effects on GnRH secretory characteristics are summarized in Figure 4}. These effects consist in a stimulatory action on GnRH pulse amplitude and a decrease in GnRH pulse frequency, as accounted again by the model results. 

Care should be taken in the mechanistic interpretation of those simulations. Direct effect in the models should be rather interpreted as acute effects than direct steroid effects on GnRH neurons, whether they are nuclear-initiated, through steroid receptors\footnote{GnRH neurons are endowed with type $\beta$ estradiol receptors \cite{ErikHrabovszky07012001}, but do not own progesterone receptors\cite{DonalCSkinner02012001}} or membrane-initiated (a recently discovered signaling way \cite{ronnekleiv_05}). But at the least, the simulation correspond to physiological short-term effect implying at most a few neuronal delays on short time scales, in contrast to the indirect, long-term effects described below.

\subsubsection{Steroid-like indirect effects}
\begin{figure}[H]
{\centering
\includegraphics[scale=0.6]{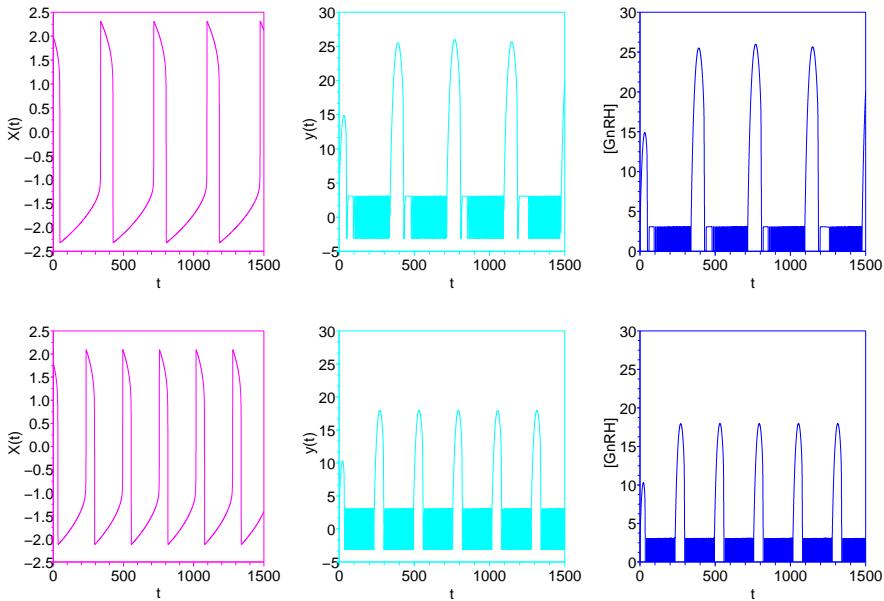}
\caption{Effect of progesterone-like priming on surge onset and amplitude}\label{effetindirectPamp}}
{\small\it Top panel: reference situation. Bottom panel situation corresponding to the absence of progesterone priming, with shortened surge onset and decreased surge amplitude. pink line: $X(t)$, cyan line: $y(t)$, blue line: GnRH output}
\end{figure}

Targeting adequate parameters in the slow system in a chronic way allow to reproduce the known effects of progesterone on the surge amplitude and delay for surge onset. 
Figures \ref{effetindirectPamp} illustrates the effect of decreasing the amplitude and increasing the frequency of the oscillations in the $X$ regulatory variable. This leads to a decrease in the delay between two consecutive GnRH surges as well as in the surge amplitude. Such combined effects mimic those that have been observed in an experimental study of the long term effect of progesterone priming \cite{AlainCaraty01011999}, which compared the GnRH surge after exposure or not to progesterone. In the absence of progesterone priming, the size of the GnRH surge was decreased and its onset shortened.

\section{Conclusion}
We addressed in this paper the question of how the GnRH generator switches from a pulsatile towards a surge secretion mode. Such a question arises on the physiological scale, when considering the behavior of average GnRH neurons whose synchronization is taken for granted. 
\noindent Zeeman {\em et al.} \cite{zeeman_03} rather tackled the question of the GnRH-induced LH surge on the pituitary gland level, considering the GnRH self-priming on gonadotroph cells as a resonance phenomenon. 

\noindent On the hypothalamic level, only the variability in the frequency of GnRH pulses (rather than its control) has been up to now the focus of mathematical models based on nonlinear dynamics \cite{brown_95,brown_94}. Our modeling approach is comparable to these previous ones in the sense that it also considers the effect of the average activity of one group of neurons on the activity of another group. But the way by which this effect is introduced differs. They used as external inputs an impulsion train, whereas we assume that both groups can be represented by the same type of equations (of FitzHugh-Nagumo type) but with different time scales. Following a 3 time-scaled approach, we have not only managed to account for the alternating pulse and surge pattern of GnRH secretion, but also for the frequency increase in the pulsatile regime. We have also unraveled the possible existence of a pause before pulsatility resumption after the surge,  which could be investigated from an experimental viewpoint. Hence the capacity of our model to display complex features interpretable against experimental evidence suggests that such a modeling approach may be a useful complement to experimental studies of neuro-endocrine systems.

\clearpage
\bibliography{manuscrit}
\bibliographystyle{siam}

\end{document}